\documentclass[]{spie}  
\usepackage[]{graphicx}
\usepackage{siunitx}
\usepackage{caption}	
\usepackage{subcaption}	

\DeclareSIUnit[per-mode=symbol]{\cms}{\cm\per\second}
\DeclareSIUnit[per-mode=symbol]{\mps}{\meter\per\second}
\DeclareSIUnit{\micron}{\micro\meter}
\title{Optimal non-circular fiber geometries for image scrambling in high-resolution spectrographs} 

\author{Julian St\"urmer\supit{a}, Christian Schwab\supit{b,c}, Stephan Grimm\supit{d}, Andr\'e Kalide\supit{d}, Adam P. Sutherland\supit{a}, Andreas Seifahrt\supit{a}, Kay Schuster\supit{d}, Jacob L. Bean\supit{a}, Andreas Quirrenbach\supit{e}
\skiplinehalf
\supit{a}University of Chicago, USA;
\supit{b}Macquarie Univ., Australia; 
\supit{c}AAO, Australia; 
\supit{d}Leibniz Institute of Photonic Technology, Germany;
\supit{e}Zentrum f\"ur Astronomie, Heidelberg
}

\authorinfo{Further author information: (Send correspondence to JS)\\ E-mail: stuermer@uchicago.edu}

\begin{document} 
  \maketitle 

\begin{abstract}
Optical fibers are a key component for high-resolution spectrographs to attain high precision in radial velocity measurements. We present a custom fiber with a novel core geometry - a 'D'-shape. From a theoretical standpoint, such a fiber should provide superior scrambling and modal noise mitigation, since unlike the commonly used circular and polygonal fiber cross sections, it shows chaotic scrambling. We report on the fabrication process of a test fiber and compare the optical properties, scrambling performance, and modal noise behavior of the D-fiber with those of common polygonal fibers. 
\end{abstract}



\section{INTRODUCTION}
\label{sec:intro}  
Optical fibers play an important role in high-precision spectroscopy. Their usage not only allows to detach the spectrograph from the telescope but fibers also offer a superior illumination stability compared to slit spectrographs, which is essential for high-precision radial velocity measurements.

In the last couple of years, astronomers have benefited from the availability of fibers with non-circular cross sections, which have been shown to provide a more stable output \cite{Avila2010,Chazelas2010} and which are used nowadays in many modern spectrographs \cite{Mahadevan2012,Stuermer2014,Furesz2014}. This experimental fact has been exploited without a detailed explanation of the theory. Often, fibers are produced for other industrial applications and might therefore not be optimized for astronomical spectrographs. In this paper we discuss the question of whether the currently used fibers can be further optimized in terms of their shape to provide higher scrambling values.

In section 2 we present simple ray trace simulations and discuss a theoretical framework that helps to predict the optical behavior of fibers with different cross sections. In section 3 we describe the manufacturing process of a D-shaped fiber that was manufactured as a model for a so called chaotic fiber. In section 4 we present experimental results of how chaotic fibers compare to other core shapes regarding their scrambling and modal noise behavior.
\section{THEORY}
Optical fibers are among the most developed and evolved optical components made out of the purest glasses available. This makes them well suited for theoretical studies and simulations. This is useful because fibers are not available in arbitrary shapes and sizes and custom fiber draws are rather expensive. 

Already in the early days of fibers being used in astronomy, light propagation in fibers and how it affects the scrambling was analyzed and the poor radial scrambling of circular fibers could be explained.\cite{Heacox1987} More recently, ray trace simulations were performed to give insight into the behavior of non circular cross section fibers.\cite{Allington-Smith2012}

\begin{figure}[htb]
\centering
\includegraphics{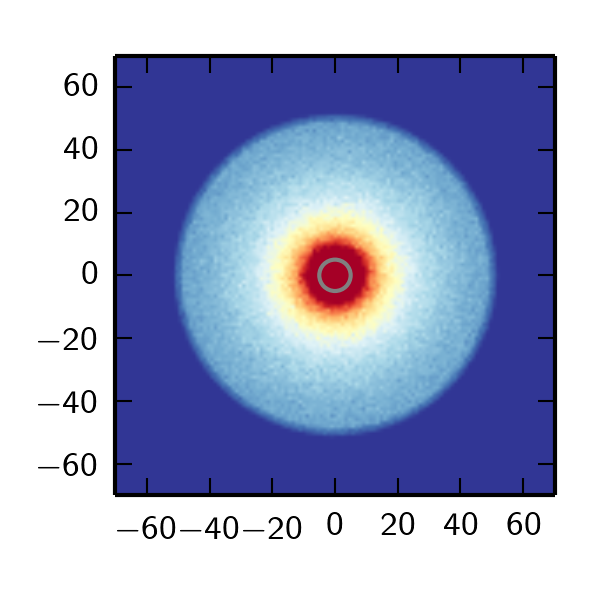}
\includegraphics{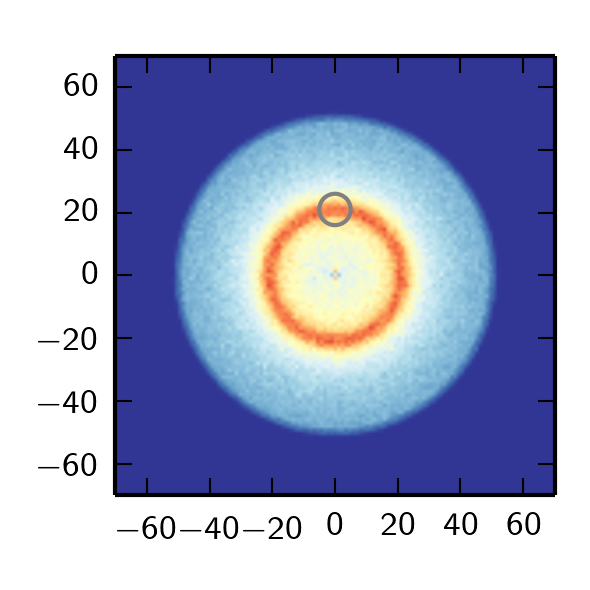}
\includegraphics{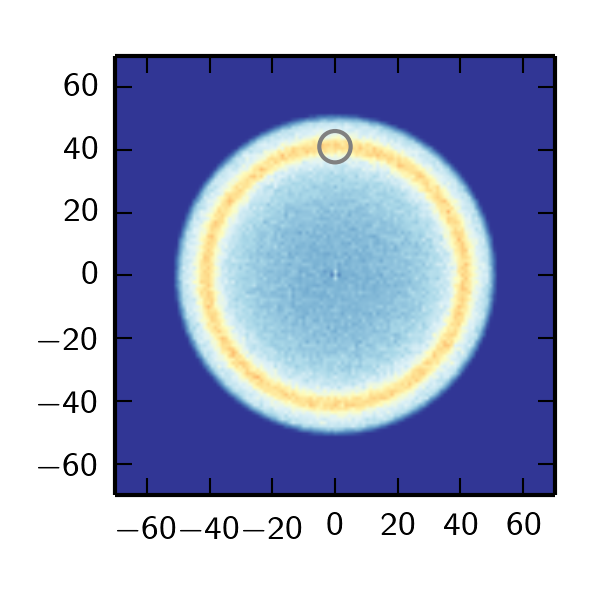}
\includegraphics{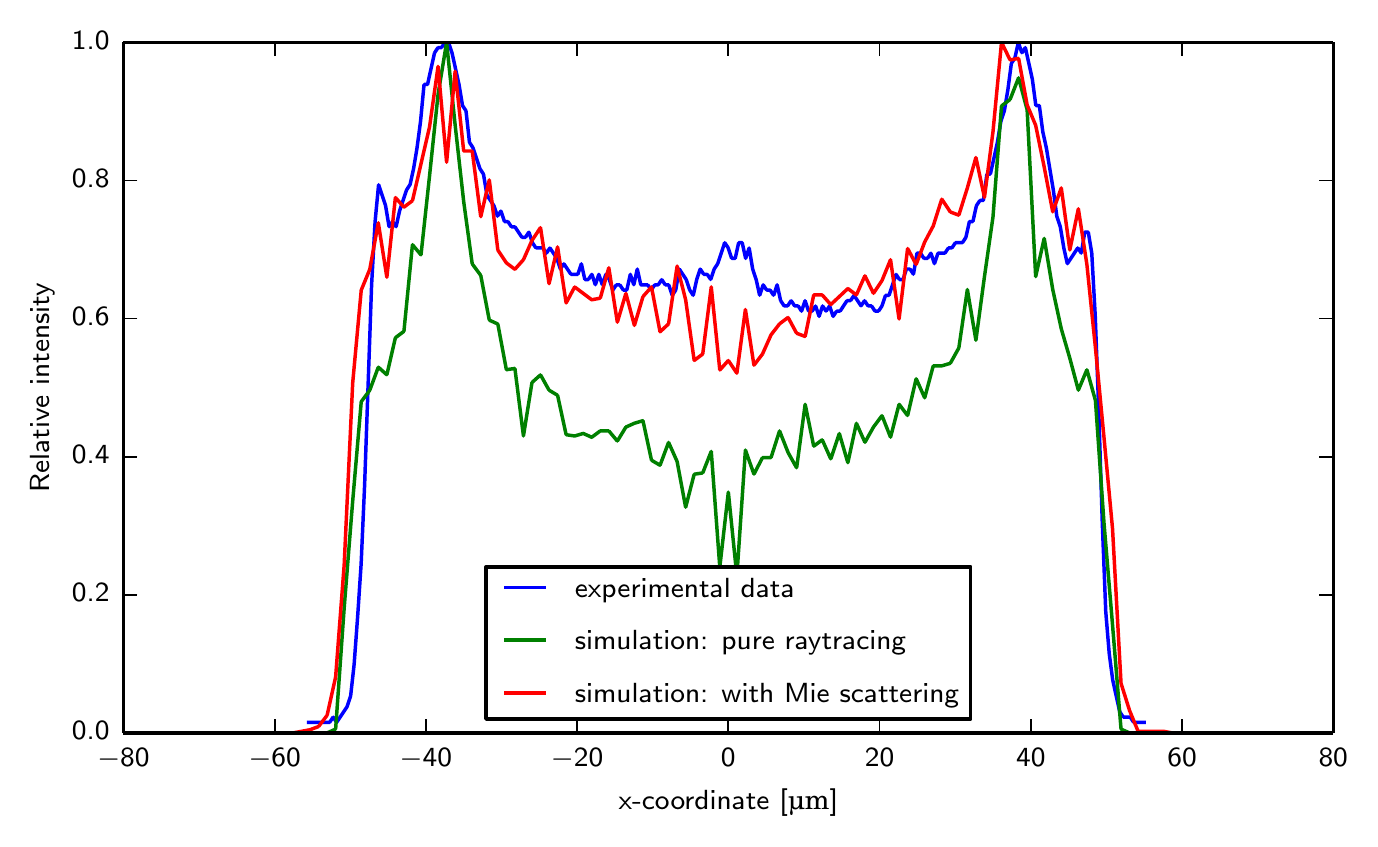}

\caption{\textbf{TOP:} Raytrace simulation of a \SI{100}{\micro\metre} circular fiber when illuminated with a \SI{5}{\micro\metre} spot. The gray circle indicates the input position of the spot. \textbf{BOTTOM:} Near field cross section of simulated and experimental data for the same fiber. By including scattering effects, a good quantitative match between experiments and simulations can be achieved.}
\label{fig:simu_circular}
\end{figure}

Here we used the commercial software ZEMAX to mimic our experiments, which are presented in section \ref{experiments}. A circular light source is imaged onto a fiber and moved across its surface. The fiber is represented by a solid glass rod surrounded by another silica tube, with their refractive indices matching the standard NA of 0.22. To validate this simple model, we compared our simulations to the measured  output of a standard \SI{100}{\micro\metre} circular fiber illuminated with a \SI{5}{\micro\metre} spot (Fig. \ref{fig:simu_circular}).

When no scattering effects are taken into account, the simulation shows more pronounced features in the near field. Using the bulk or end-face scattering functions of ZEMAX, a good match between the experiment and the simulation can be achieved (see bottom of Fig. \ref{fig:simu_circular}). The scattering parameters where set manually to match the experimental results. No attempt was made to provide physical justification and trace different scattering processes. However, providing empirical data of the scattering properties of fibers in ZEMAX and accounting for them accordingly might allow for quantitative models. This will be necessary for analyzing FRD effects, as scattering effects play an important role here.\cite{Haynes2011} In general, random scattering effects will decrease the contrast of illumination inhomogeneities.

By assuming the large core optical fiber to be homogeneous and ideal in the Z-direction, the fiber model can be further simplified to a 2D model where the ray propagation is modeled by the friction-less motion of a particle within the fiber core's boundary. This system is called a \textit{dynamical billiard}, a standard problem of dynamical system theory. The equivalence of a step index fiber with a dynamical billiard can also be derived directly from the Helmholtz equation.\cite{Doya2002} 
The basic idea is that the underlying dynamics of the billiard system tells us more about the optical behavior and that scrambling or modal properties can be derived from this theory.

The dynamics in a billiard can show all types of behavior from regular motion to chaotic systems, depending exclusively on the shape of the boundary. In numerous studies, the dynamic behavior of all kinds of boundary shapes has been extensively analyzed. The findings that are interesting in the context of scrambling are that circular and elliptical billiards are \textit{integrable}. In an integrable billiard, the motion for all starting conditions is regular. As neighboring starting conditions lead to almost identical motion patterns in such a system, there is a strong coupling between the input and the output in fibers with such a shape.

Polygonal shapes are special, because their boundary is not twice differentiable and they are called \textit{pseudo-integrable}.\cite{Richens1981} Two properties of polygons are of special interest for us: First, it can be shown that all trajectories uniformly cover the whole cross section unless the trajectory is a (closed) periodic orbit.\cite{DeMarco2011} Second, the dynamics are not chaotic, meaning that neighboring rays do not diverge exponentially.
Slight perturbations will propagate, but perturbations do not grow or diminish.
While the first point is favorable in terms of illumination homogeneity and stability, the second means that neighboring trajectories are still correlated. 

In contrast, in a chaotic billiard, neighboring trajectories quickly diverge. After a few bounces it is impossible to trace a trajectory back to its exact initial launching conditions. Therefore, in this idealized picture, a fiber with underlying chaotic dynamics is a perfect scrambler, as it 'loses memory' of its initial state.

\begin{figure}
\centering
\includegraphics[width=0.32\textwidth]{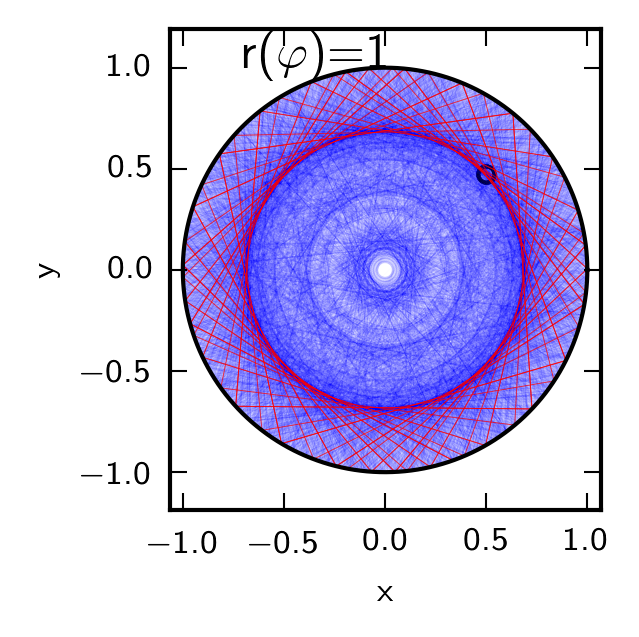}
\includegraphics[width=0.32\textwidth]{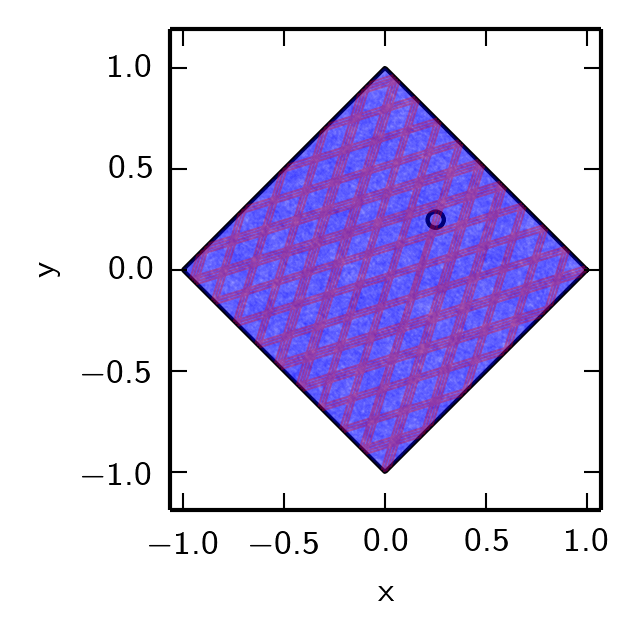}
\includegraphics[width=0.32\textwidth]{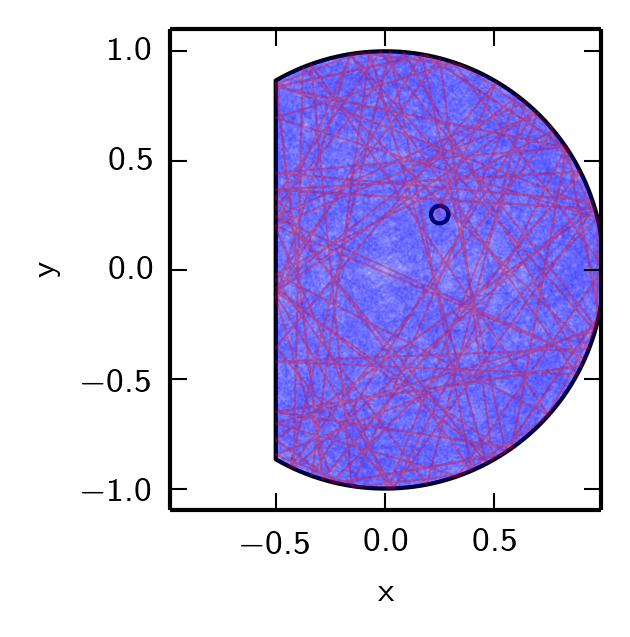}
\caption{Three different billiard tables illustrating the different dynamical behavior. The black circle indicates the initial starting positions of 200 trajectories (blue) with random orientation. In red an arbitrary ray is highlighted. The circular billiard shows regular motion for all initial conditions. The highlighted trajectory shows a clear visible caustic - a boundary that will never be crossed by this orbit. For the square billiard, all orbits cover uniformly the table but the trajectories still follow a regular pattern. For the D-shape the trajectories are chaotic, no regular pattern can be observed and orbits with similar initial conditions quickly diverge.}
\label{fig:billiards}
\end{figure}

For an illustration of the different dynamical behavior of the mentioned boundary shapes see Fig.\ref{fig:billiards}.

It should be noted that the concepts presented here in a simplified, pure ray-based picture also hold true when fibers are treated as wave-guides where modal effects complicate the situation. In fact, it is has been shown theoretically and experimentally that the modal statistics in a chaotic fiber are different from those in integrable systems.\cite{Doya2002}

\section{A CHAOTIC FIBER}
Based on the conclusions from the previous section and considering the results of other groups in the context of fiber lasers\cite{Doya2002, Michel2009}, we opted for a fiber shape that shows chaotic dynamics. We have chosen a D-shape, where the straight cut occurs at 3/4 of the fiber diameter. D-shaped fibers have already been investigated experimentally as well as theoretically in other studies.\cite{Ree1999, Doya2002} This shape is also relatively easy to manufacture. For a spectrograph, however, other (chaotic) shapes are better suited: stadium shapes (half-circles connected by straight sections) or regular polygons with rounded corners better fit a spectrograph slit.

\begin{figure}[hbt]
\centering
\includegraphics[width=0.49\textwidth]{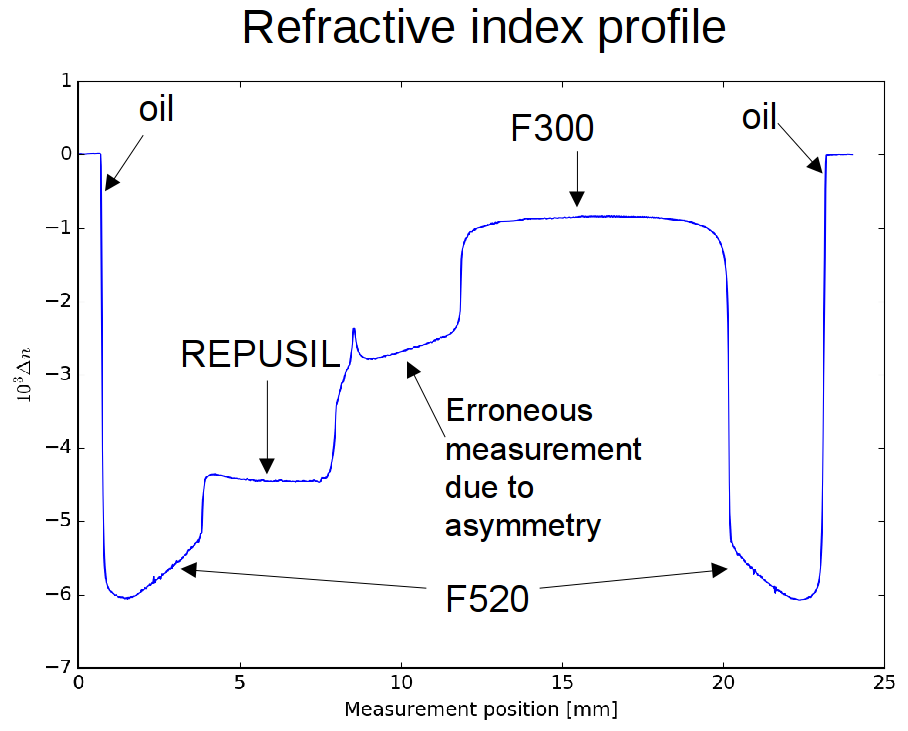}
\includegraphics[width=0.49\textwidth]{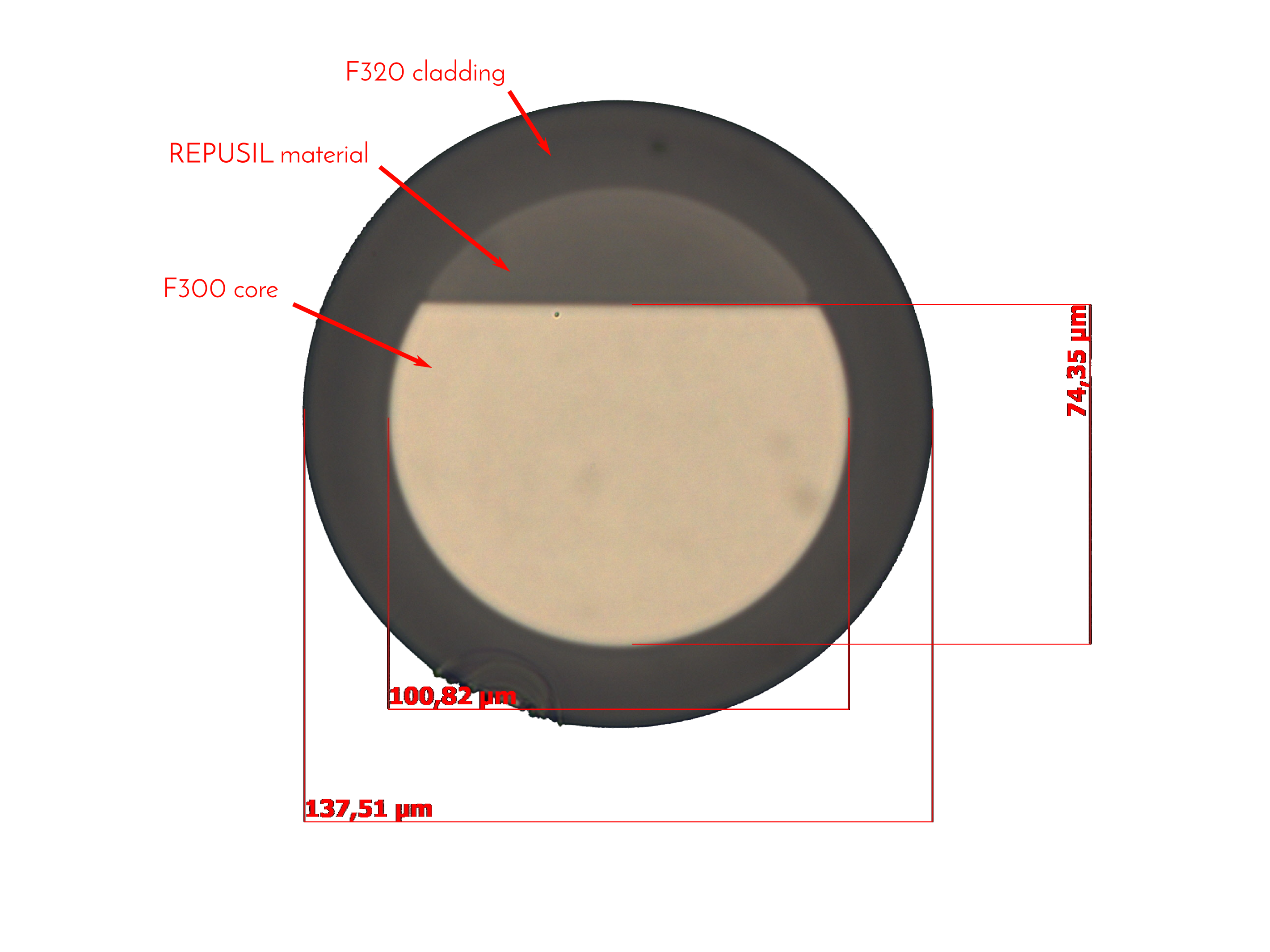}
\caption{\textbf{LEFT:} Measured refractive index profile of the preform. \textbf{Right:} Microscope image of a cleaved \SI{100}{\micro\metre} D-shaped fiber. Due to a slight mismatch of the refractive index of the REPUSIL material and the cladding, some light is guided in REPUSIL part when illuminated, making it appear brighter than the cladding. }
\label{fig:dshape}
\end{figure}

Our fiber was manufactured by IPHT Jena. The core of the fiber is made of a synthetic silica (F300, Heraeus) rod with the D-cut mechanically ground. The core was inserted into a commercial F-doped tube (F320, Heraeus) with dimensions of $22.5 \times 16.4$\, mm (OD x ID). The gap between the cladding and the D-shaped core was filled with a refractive index adjusted F-doped silica glass fabricated by the reactive powder sintering technology (a.k.a. REPUSIL process).\cite{Schuster2014}
Finally, fibers with \SI{35}{\micro\metre}, \SI{67}{\micro\metre} and \SI{100}{\micro\metre} cores, each \SI{100}{\metre} long, were drawn from the preform and coated with Ormocer. Figure \ref{fig:dshape} shows the refractive index cross-section of the preform and a visual cross-section of the \SI{100}{\micro\metre} fiber. Due to a slight mismatch in the refractive index between the REPUSIL material and the outer cladding, some light is guided in the missing D-section, but only when directly illuminated. \\
Compared to a commercial OptranWF fiber from CeramOptec, we measured a \SI{10}{\percent} reduced throughput from \SIrange{450}{550}{\nano\metre} for an equally long fiber. For longer wavelengths, the difference in throughput decreases (to \SI{6}{\percent} at \SI{850}{\nano\metre}). We therefore suspect that scattering effects are the dominant source of the increased light loss.

\section{SCRAMBLING AND MODAL NOISE}
\label{experiments}
\begin{figure}
\centering
\includegraphics[width=0.49\textwidth]{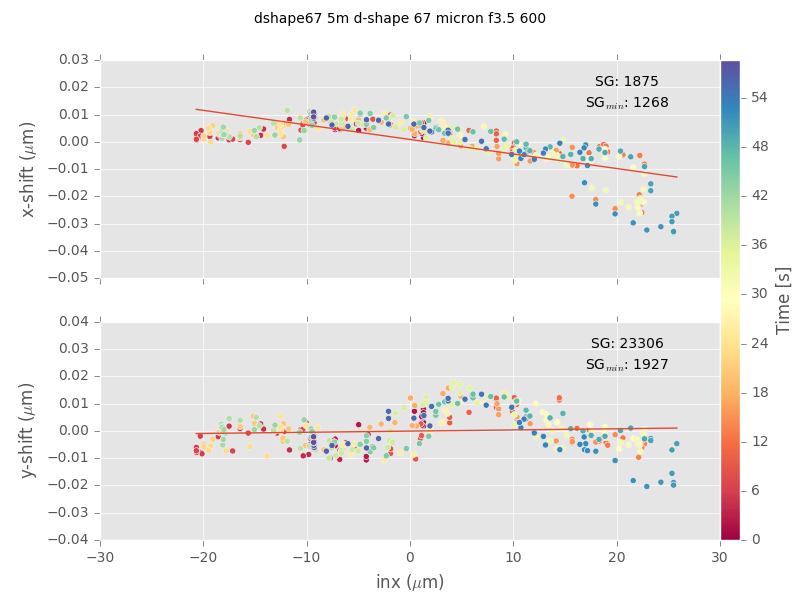}
\includegraphics[width=0.49\textwidth]{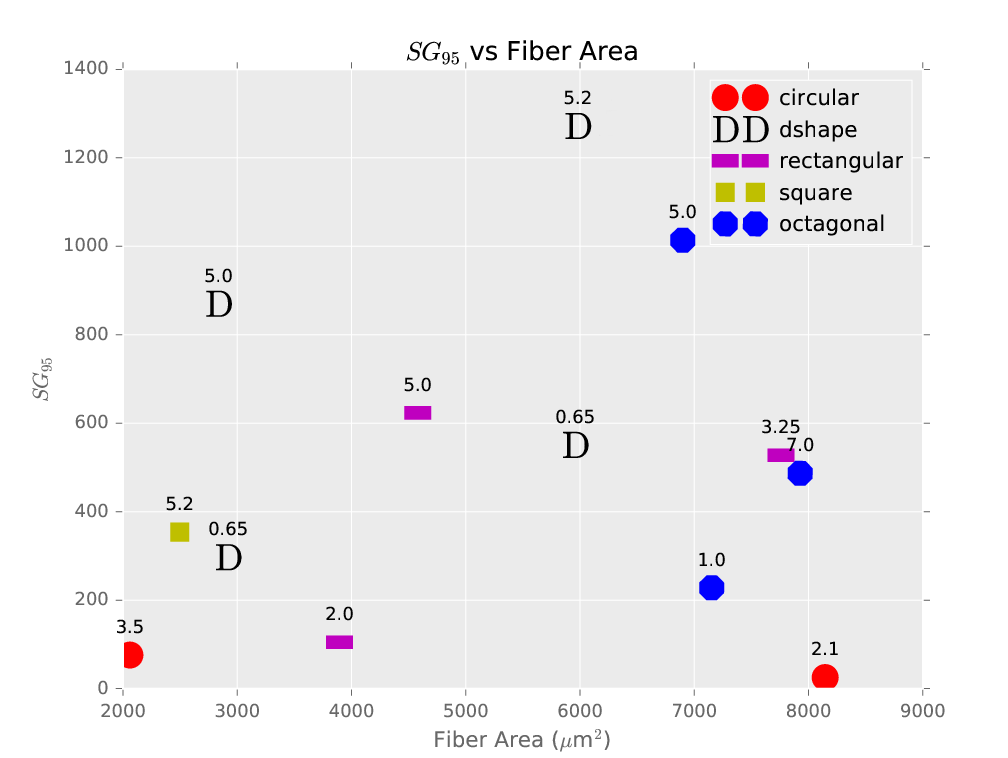}
\caption{\textbf{Left:} Scrambling gain measurement for \SI{67}{\micron} d-shaped fiber at $f$/3.5 and $\lambda$=\SI[separate-uncertainty=true]{600+-20}{\nano\meter}. The classical, linear scrambling gain is shown as a linear fit, while the $SG_{min}$ metric is derived from the full distribution of datapoints. The image barycenter of the output fiber face was measured in X (top) and Y (bottom) while sweeping the location of a \SI{10}{\micron} pinhole multiple times across the input face of the fiber. \textbf{Right:}
Scrambling gain vs fiber area in $\mu$m$^2$ for all tested fibers, averaged over three wavelengths and for $f$/3.5 and $f$/4. The  length of each fiber in meters is shown above each marker.}
\label{fig:scrambling}
\end{figure}

To measure the scrambling gain of the fibers, a small pinhole is imaged onto the front face of the fiber and moved across it. Output and input are recorded and analyzed. The setup as well as the data reduction is described in detail in another conference proceeding. \cite{adam} The pinhole is swept across the fiber input several times. A typical result can be seen in Fig. \ref{fig:scrambling}. We calculate the scrambling gain as 
\begin{equation}
SG_{min} = \Delta d_{max} / \Delta F_{max},
\end{equation}
where $\Delta d_{max}$ is the maximum movement of the pinhole. $\Delta F_{max}$ is the maximum shift of the barycenter of the near field at the output of the fiber. We discard the top and bottom \SI{2.5}{\percent} of the distribution in order not to be dominated by outliers. This modification of the scrambling gain leads to more reproducible results, as fibers often do not show a strict linear relation between input and output shifts.

Fig. \ref{fig:scrambling} shows the results of scrambling gain measurements for various fibers available in our lab. 
As expected, circular fibers show the lowest SG. In general the length of the fiber also matters. The number above the markers in Fig. \ref{fig:scrambling} shows the length in meters and one can see that SG increases significantly with length for the same fiber type.
The D-shaped fiber outperforms other fibers of similar size and length - a strong indication for superior scrambling properties. Each data point in Fig. \ref{fig:scrambling} is an average over many scrambling measurements taken with different input $f$-ratios and different broadband ($\sim$\SI{40}{\nano\metre} bandwidth) filters from \SIrange{400}{900}{\nano\metre}.

For measuring the influence of the fiber shape on the modal behavior of the fiber, we replaced the white light illumination source with a HeNe laser for three different fibers, a circular, an octagonal, and a d-shaped fiber, each with \SI{100}{\micron} diameter. In a first test we calculated the Michelson contrast in the near-field of each fiber. The fiber is left untouched during the measurement. After that the fiber is agitated by attaching a magnet onto the protection tube of the fiber and shaking it by the varying magnetic field of an electric coil at \SI{60}{\Hz}. The fibers were bound together to assure a common mechanical movement. 

\begin{figure}
\centering
\includegraphics[width=0.32\textwidth]{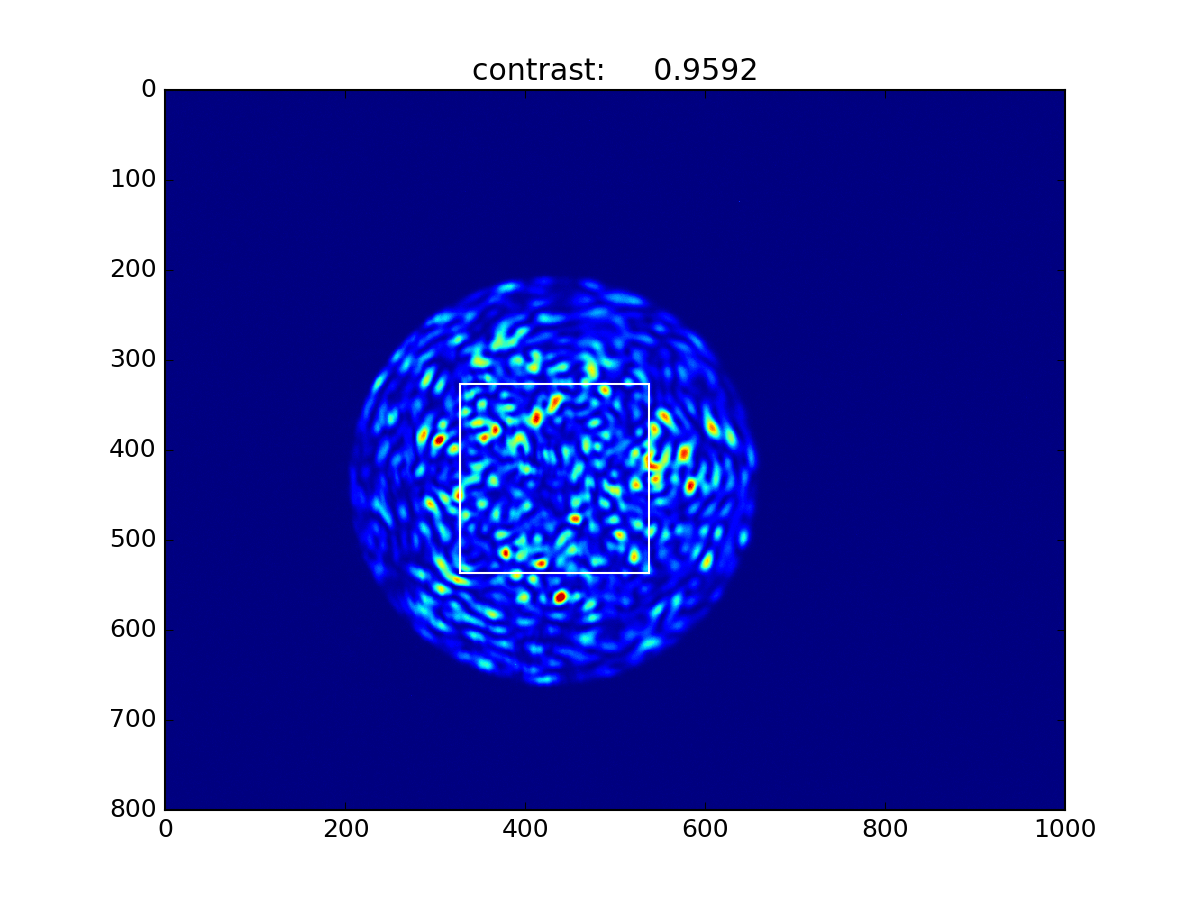}
\includegraphics[width=0.32\textwidth]{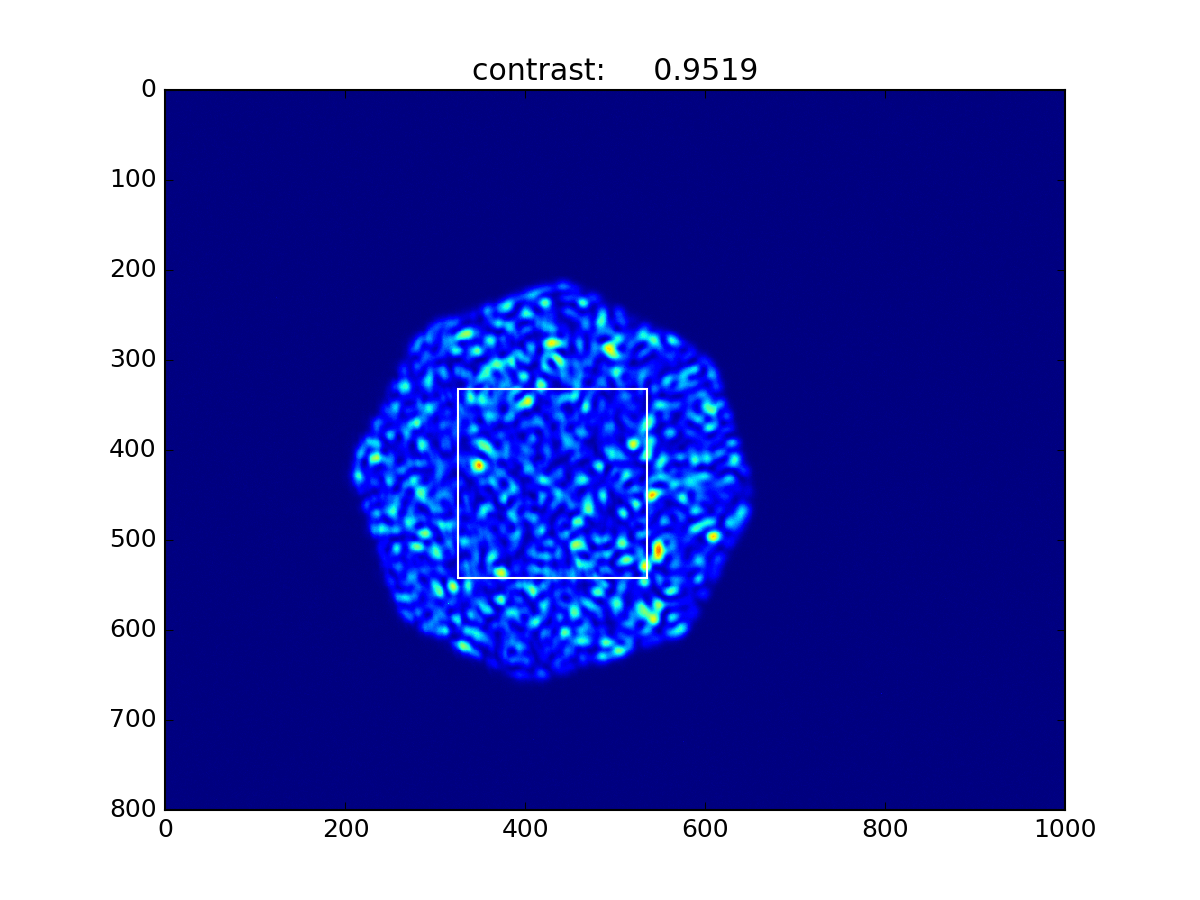}
\includegraphics[width=0.32\textwidth]{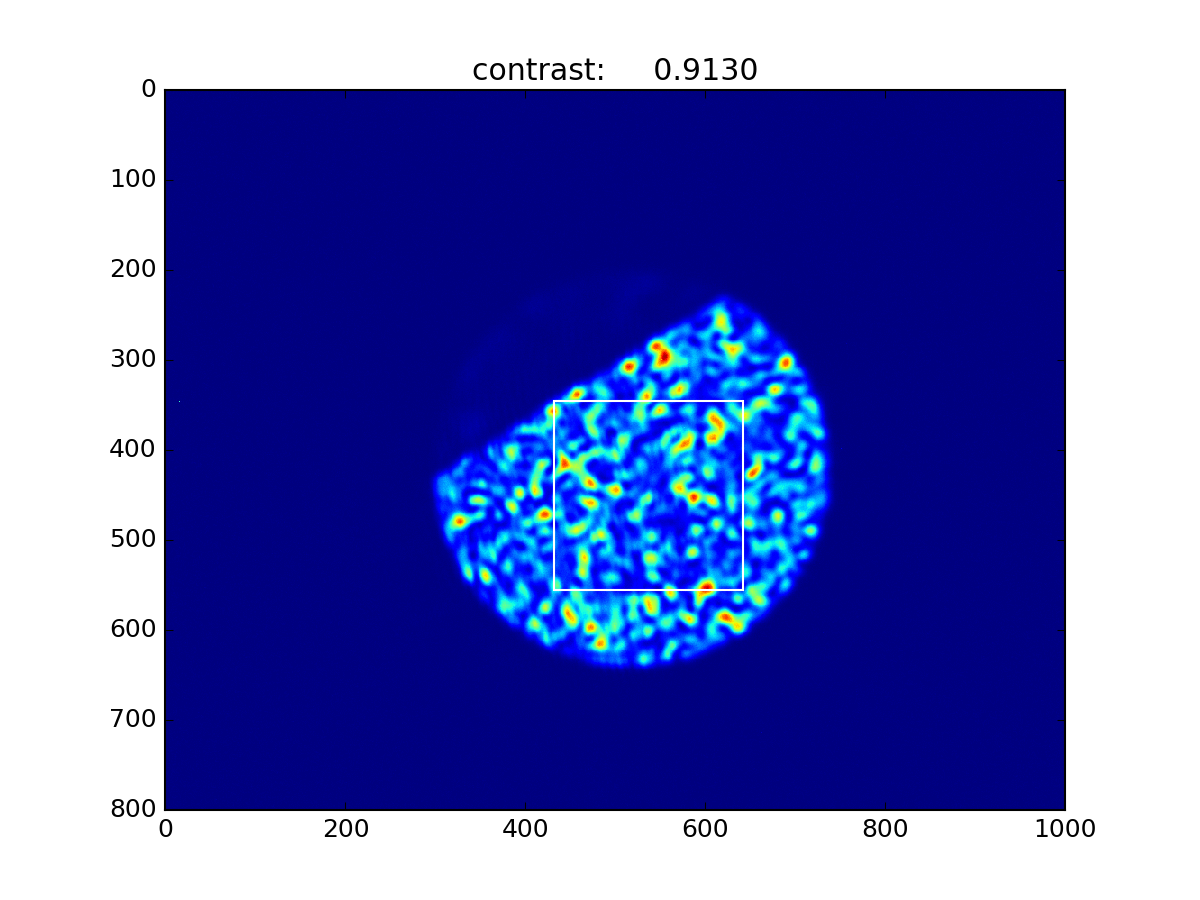}
\includegraphics[width=0.32\textwidth]{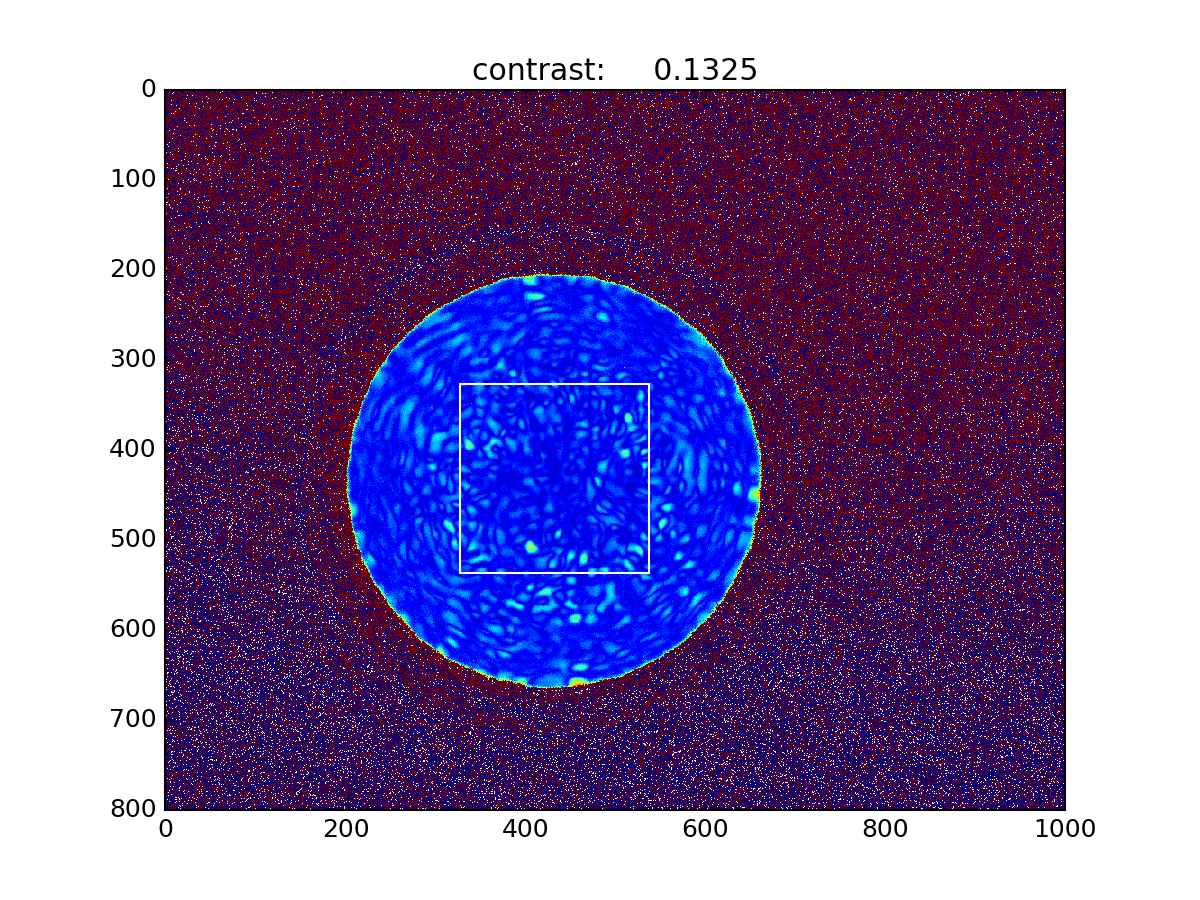}
\includegraphics[width=0.32\textwidth]{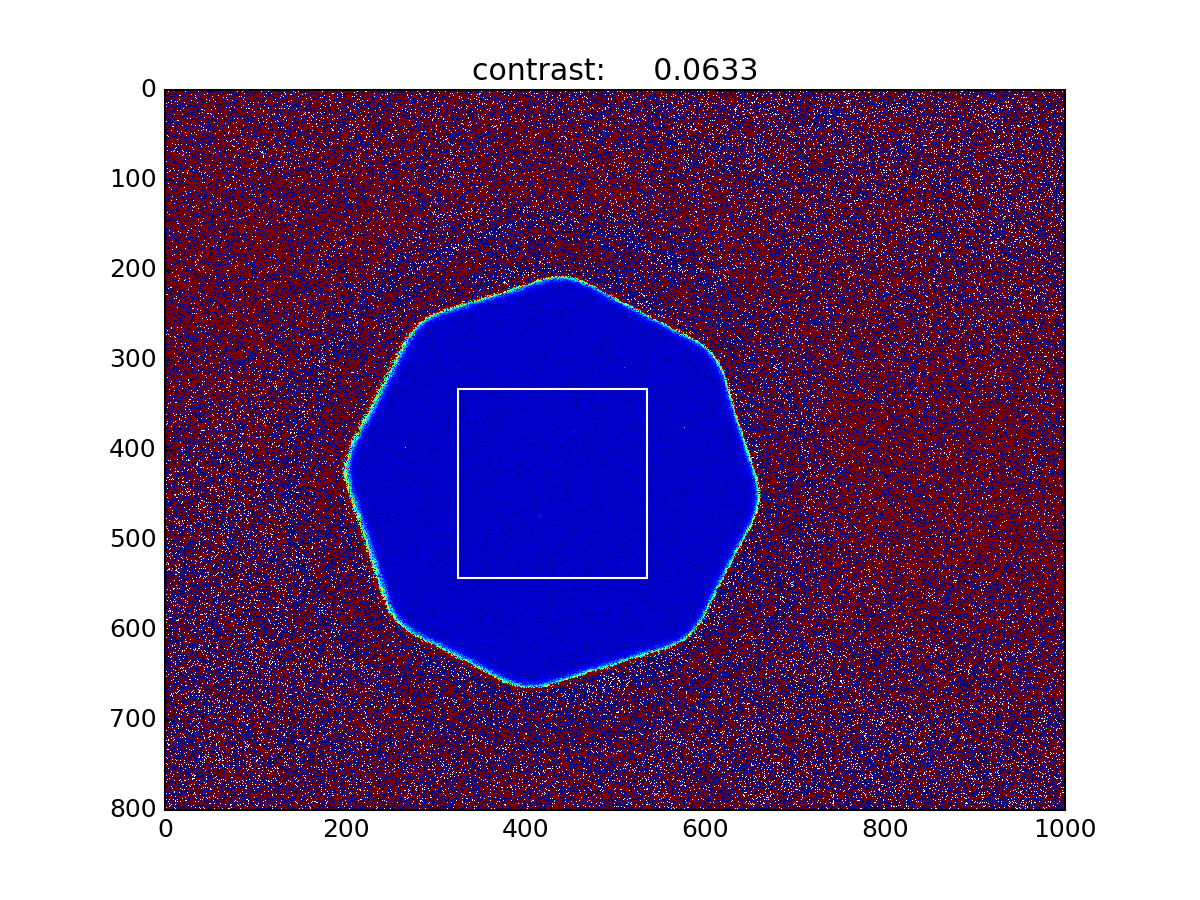}
\includegraphics[width=0.32\textwidth]{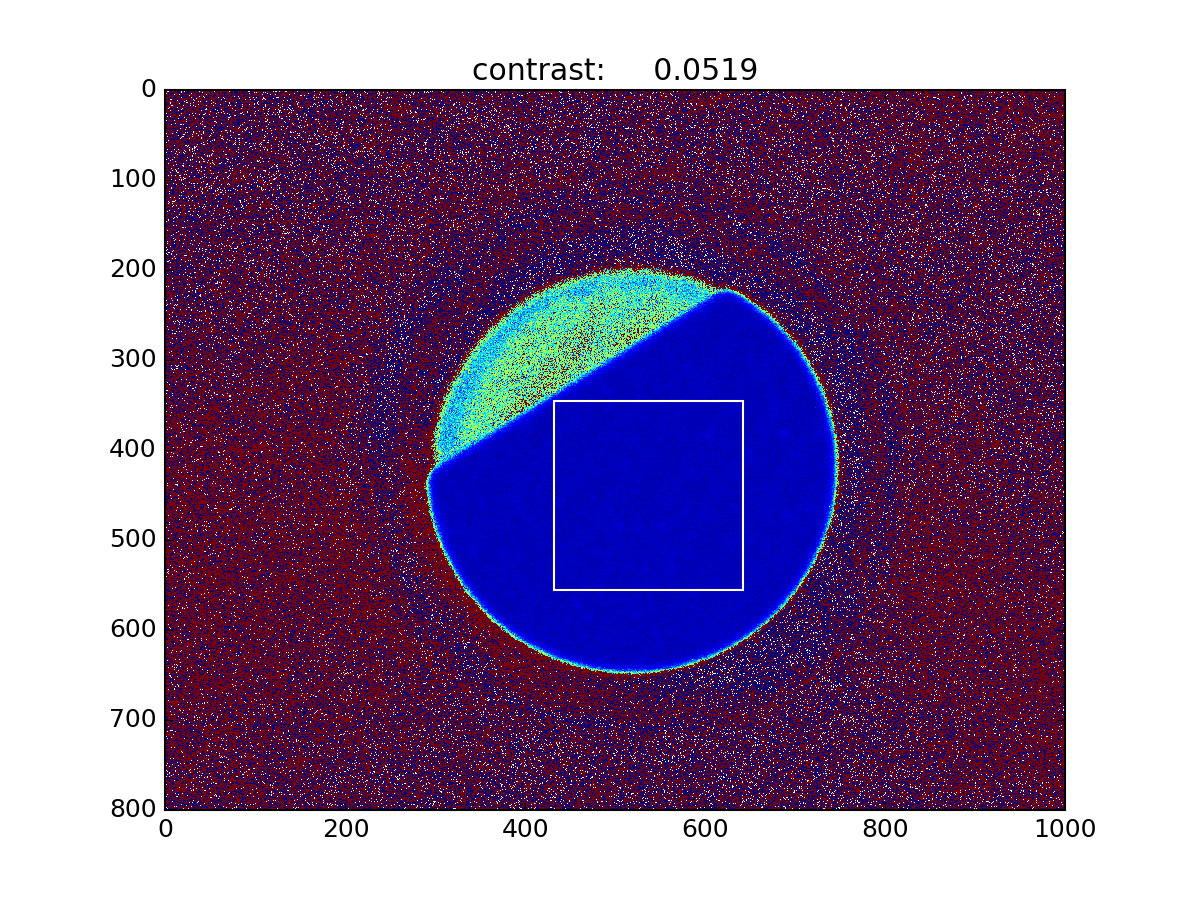}

\caption{\textbf{TOP ROW:} Speckle images of unshaken fibers under monochromatic illumination show a Michelson contrast close to 1 regardless of their shape.
\textbf{BOTTOM ROW:} Shaking the fibers reduces speckle contrast. Circular fibers perform worst (Michelson contrast of 0.13) compared to octagonal and d-shaped fibers (Michelson contrast of 0.06 and 0.05, respectively), as their speckle patterns change less for the same mechanical agitation.}
\label{fig:modalnoise}
\end{figure}

For the static fiber, the Michelson contrast is close to 1, as expected for highly coherent light sources (see Fig. \ref{fig:modalnoise}). Here, the shape of the fiber makes no difference, which also agrees with theory. 
To calculate the Michelson contrast for the agitated fiber, we averaged 21 \SI{1}{\second} frames each. Agitation reduces the contrast in the near-field of the circular fiber to about 0.13. For the octagonal and the D-shape fiber we find values which are smaller by a factor of two. We conclude that modal suppression by agitation in a circular fiber is less efficient. The speckle pattern is in a sense more stable and less affected by perturbations. Further measurements have to show if there is a significant difference between the speckle behavior of polygonal fibers versus the D-shaped fiber.

\begin{figure}
\centering
\includegraphics[width=0.80\textwidth]{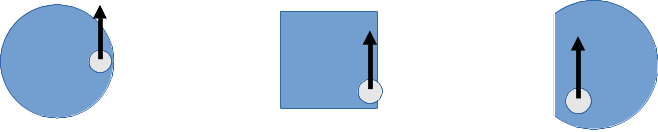}
\includegraphics[width=0.32\textwidth]{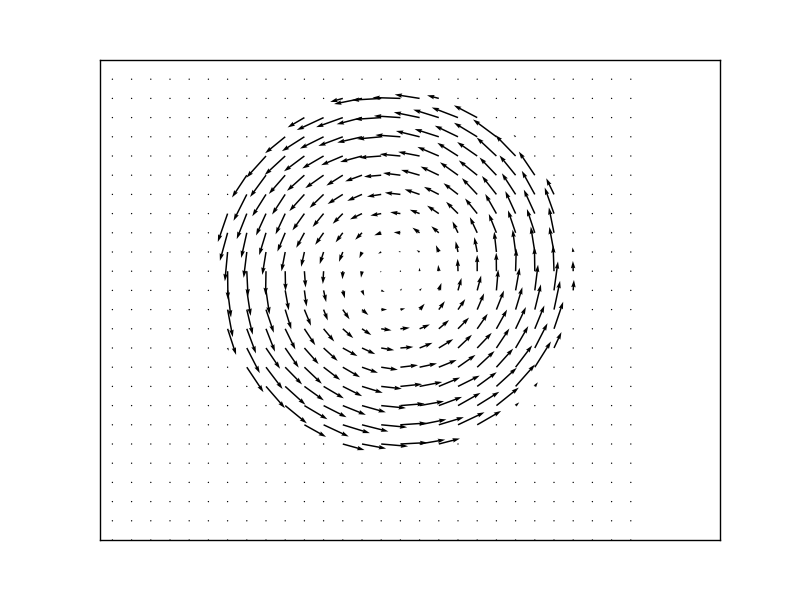}
\includegraphics[width=0.32\textwidth]{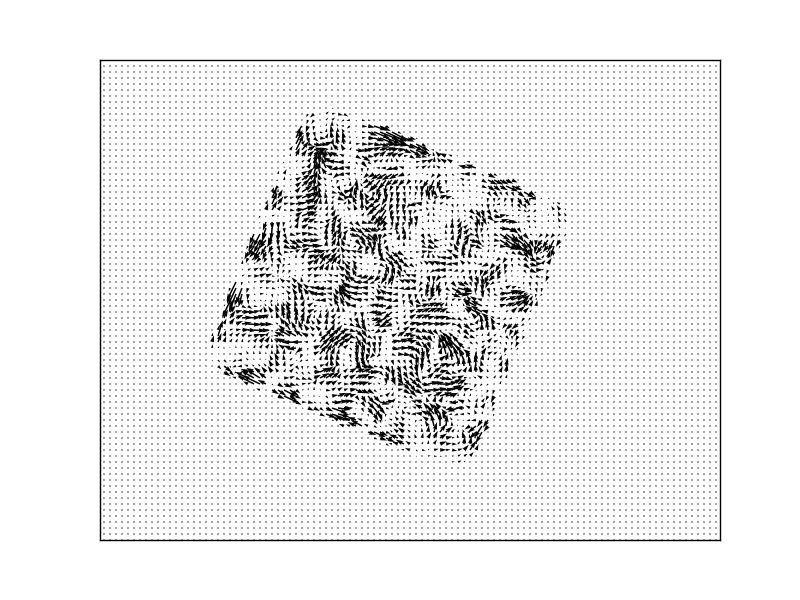}
\includegraphics[width=0.32\textwidth]{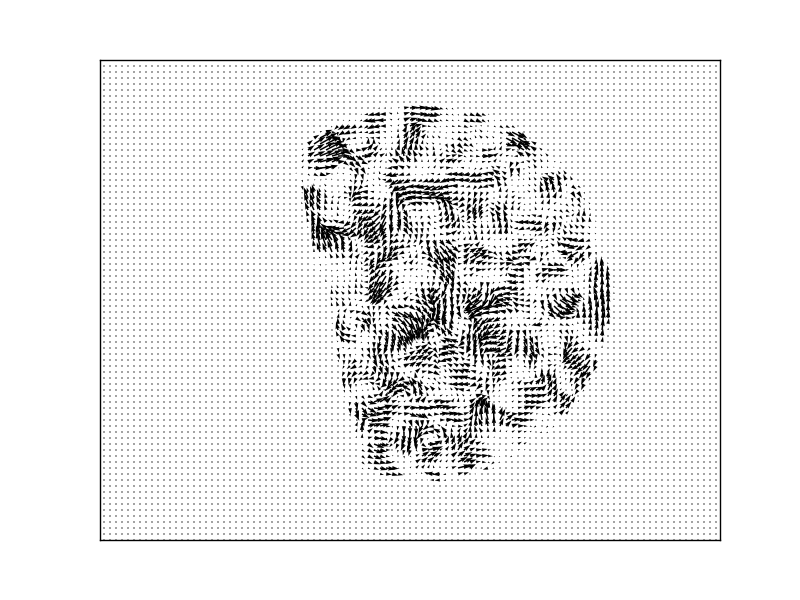}
\caption{\textbf{TOP ROW:} Illustration of the fiber input: a small laser spot is slowly moved tangentially to the fiber boundary. For each measurement about 100 images are recorded.
\textbf{BOTTOM ROW:} Calculated optical flow from the resulting near field speckle pattern images. The arrows indicate the direction and speed of how the speckles moved on average. While the circular fiber shows a highly correlated movement of the speckles, the other shapes show no visible large scale structures.}
\label{fig:opticalflow}
\end{figure}

More direct proof of the highly correlated movement of speckles in a circular fiber is given by analyzing a timeseries of speckle images of each fiber. Instead of agitating the fiber, the laser spot is moved across the fiber input face, similar to a scrambling gain measurement but with a drastically reduced scanning speed. From the resulting series of speckle images (around 100 per fiber), the optical flow is calculated for all consecutive images. Figure \ref{fig:opticalflow} shows the averaged optical flow across the fiber. The circular fiber shows a strongly correlated movement of the speckles (i.e. a rotation-like movement) when moving tangential to the boundary. This cannot be seen for the square fiber or the D-shaped fiber. This may be seen as a direct indication of how the speckle movement reflect the underlying dynamics of the system.
\section{CONCLUSION AND OUTLOOK}
We have presented ray tracing simulations and a theoretical approach to better understand the different performance in scrambling and modal noise suppression of differently shaped optical fibers.
Based on the conclusion that fibers with chaotic dynamics should perform best, we manufactured D-shaped fibers and tested their scrambling and modal behavior. First results show excellent scrambling performance, possibly outperforming polygonal fibers. Further measurements are planned to investigate the significance of these results. Similarly, the modal noise suppression in such fibers is at least as good as in polygonal fibers. Systematic experiments to measure the speckle statistics are planned in order to check for similarities and differences in the speckle statistics of polygonal fibers and the D-shaped fiber.

\acknowledgments
The authors thank Stephan Meister for a fruitful discussion regarding the optical flow measurements.

\bibliography{report2}   

\begin{thebibliography}{10}

\bibitem{Avila2010}
Avila, G., Singh, P., and Chazelas, B., ``{Results on fibre scrambling for high
  accuracy radial velocity measurements},'' {\em Proc. SPIE}~{\bf 7735},
  773588--773588--9 (2010).

\bibitem{Chazelas2010}
Chazelas, B., Pepe, F., Wildi, F., Bouchy, F., Perruchot, S., and Avila, G.,
  ``{New scramblers for precision radial velocity: square and octagonal
  fibers},'' {\em Proc. SPIE}~{\bf 7739},  773947--773947--9 (2010).

\bibitem{Mahadevan2012}
Mahadevan, S., Ramsey, L., Bender, C., Terrien, R., Wright, J.~T., Halverson,
  S., Hearty, F., Nelson, M., Burton, A., Redman, S., Osterman, S., Diddams,
  S., Kasting, J., Endl, M., and Deshpande, R., ``{The Habitable-Zone Planet
  Finder: A Stabilized Fiber-Fed NIR Spectrograph for the Hobby-Eberly
  Telescope},'' {\em SPIE 2012 Astronomical Instrumentation and Telescopes
  conference} ,  14 (2012).

\bibitem{Stuermer2014}
St{\"{u}}rmer, J., Stahl, O., Schwab, C., Seifert, W., Quirrenbach, A., Amado,
  P.~J., Ribas, I., Reiners, A., and Caballero, J.~A., ``{CARMENES in SPIE
  2014. Building a fibre link for CARMENES},'' {\em Proc. SPIE}~{\bf 9151},
  915152--915152--12 (2014).

\bibitem{Furesz2014}
Fűr{\'{e}}sz, G., Epps, H., Barnes, S., Podgorski, W., Szentgyorgyi, A.,
  Mueller, M., Baldwin, D., Bean, J., Bergner, H., Chun, M.-Y., Crane, J.,
  Evans, J., Evans, I., Foster, J., Gauron, T., Guzman, D., Hertz, E.,
  Jord{\'{a}}n, A., Kim, K.-M., McCracken, K., Norton, T., Ordway, M., Park,
  C., Park, S., Plummer, D., Uomoto, A., and Yuk, I.-S., ``{The G-CLEF
  spectrograph optical design},'' {\em Proc. SPIE}~{\bf 9147},
  91479G--91479G--9 (2014).

\bibitem{Heacox1987}
Heacox, W.~D., ``{Radial image transfer by cylindrical, step-index optical
  waveguides},'' {\em Journal of the Optical Society of America A}~{\bf 4},
  488--493 (Mar. 1987).

\bibitem{Allington-Smith2012}
Allington-Smith, J., Murray, G., and Lemke, U., ``{Simulation of complex
  phenomena in optical fibres},'' {\em Monthly Notices of the Royal
  Astronomical Society}~{\bf 427}(2),  919--933 (2012).

\bibitem{Haynes2011}
Haynes, D.~M., Withford, M.~J., Dawes, J.~M., Lawrence, J.~S., and Haynes, R.,
  ``{Relative contributions of scattering, diffraction and modal diffusion to
  focal ratio degradation in optical fibres},'' {\em Monthly Notices of the
  Royal Astronomical Society}~{\bf 414}(1),  253--263 (2011).

\bibitem{Doya2002}
{Doya}, V., {Legrand}, O., {Mortessagne}, F., and {Miniatura}, C., ``{Speckle
  statistics in a chaotic multimode fiber},'' {\em Physical Review E}~{\bf 65},
   056223 (May 2002).

\bibitem{Richens1981}
{Richens}, P.~J. and {Berry}, M.~V., ``{Pseudointegrable systems in classical
  and quantum mechanics},'' {\em Physica D Nonlinear Phenomena}~{\bf 2},
  495--512 (June 1981).

\bibitem{DeMarco2011}
DeMarco, L., ``{The conformal geometry of billiards},'' {\em Bulletin of the
  American Mathematical Society}~{\bf 48},  33–33 (Jan. 2011).

\bibitem{Michel2009}
Michel, C., Tascu, S., Doya, V., Legrand, O., and Mortessagne, F.,
  ``{Gain-controlled wave chaos in a chaotic optical fibre},'' {\em Journal of
  the European Optical Society - Rapid publications}~{\bf 4}(0) (2009).

\bibitem{Ree1999}
{Ree}, S. and {Reichl}, L.~E., ``{Classical and quantum chaos in a circular
  billiard with a straight cut},'' {\em Physical Review E}~{\bf 60},
  1607--1615 (Aug. 1999).

\bibitem{Schuster2014}
{Schuster}, K., {Unger}, S., {Aichele}, C., {Lindner}, F., {Grimm}, S.,
  {Litzkendorf}, D., {Kobelke}, J., {Bierlich}, J., {Wondraczek}, K., and
  {Bartelt}, H., ``{Material and technology trends in fiber optics},'' {\em
  Advanced Optical Technologies}~{\bf 3},  447--468 (Aug. 2014).

\bibitem{adam}
{Sutherland}, A., {St\"urmer}, J., {Miller}, K., {Seifahrt}, A., and {Bean},
  J.~L., ``{Characterizing octagonal and rectangular fibers for MAROON-X},'' in
  [{\em Advances in Optical and Mechanical Technologies for Telescopes and
  Instrumentation}{\nolinebreak\hspace{0.1em}]},  {\em Proc.~SPIE} {\bf 9912},
  9912185 (June 2016).

\end{thebibliography}
\bibliographystyle{spiebib}   

\end{document}